\documentclass[preprint,showpacs,preprintnumbers,amsmath,amssymb]{revtex4}
\usepackage{graphicx}% Include figure files
\usepackage{rotating}
\usepackage{epsf}
\usepackage{dcolumn}% Align table columns on decimal point
\begin{document}
\preprint{preprint - thermoelectric group, iitp/iitb}
\title{Extreme Sensitivity of Magnetic Properties on the Synthesis Routes in La$_{0.7}$Sr$_{0.3}$MnO$_3$}
\author{Ashutosh Kumar$^{1}$, Himanshu Sharma$^{2}$, C. V. Tomy$^{2}$, Ajay D. Thakur$^{1,\,3}$}
\affiliation{$^{1}$Department of Physics, Indian Institute of Technology Patna, Patna 801118, India \\
$^{2}$Department of Physics, Indian Institute of Technology Bombay, Mumbai 400076, India 
\\
$^{3}$Center for Energy and Environment, Indian Institute of Technology Patna, Patna 801118, India}
\date{\today}
\begin{abstract}
La$_{0.7}$Sr$_{0.3}$MnO$_3$ polycrystalline samples have been prepared using different synthesis routes. X-ray Diffraction (XRD) confirms that the samples are of single phase with R$\bar{3}$c space group. The surface morphology and particle size has been measured using Field Emission Scanning Electron Microscopy (FESEM). Magnetic measurement shows that the magnetization in the materials are affected by low crystallite size which destroys the spin ordering due to strain at grain boundaries and in turn also lead to reduction in magnetization as well as an enhanced coercivity in the material. 
\end{abstract}
\maketitle
\section{Introduction}
Magnetic measurements of divalent element substituted rare earth manganites, R$_{1-x}$A$_x$MnO$_3$ where R= rare earth element (La, Pr, Sm etc.), $A=$ divalent alkaline element (Sr, Ca, Ba, etc.), having perovskite structure have been widely investigated \cite{ref1}. The interesting physical properties in these materials like colossal magnetoresistance, metal-insulator transition, etc., where interplay of spin, charge and orbital coupling leads to massive interest due to many application in spintronics devices. However, The very discovery of phenomenon like double exchange, Jahn-Teller distortion have been successful to explain the physics behind the magnetic properties of these materials, where charge and spin ordering leads to conductivity and ferromagnetism at certain substitution at A-site by alkaline elements (following Hund’s rule).  It has been widely investigated\cite{ref1} that when La site in LaMnO$_3$ is partially replaced by Sr, it becomes conducting and ferromagnetic i.e. La$_{0.7}$Sr$_{0.3}$MnO$_3$. The magnetic properties in La$_{0.7}$Sr$_{0.3}$MnO$_3$ are affected by geometrical factors like shape, size, and structure. Since ferromagnetic materials consist of domains which are spontaneously magnetized separated by domain walls, but formation of domains and domain walls depends on different kinds of energy and also these domains and domains wall can be affected due to size of the crystallite present in the material. In this report, we have investigated the crystallite size affect on magnetic properties of La$_{0.7}$Sr$_{0.3}$MnO$_3$. 

\section{Experimental Details}
La$_{0.7}$Sr$_{0.3}$MnO$_3$ (LSMO) samples were prepared by sol gel (SG) route, standard solid state route (SSR) and ball milling (BM) method. In standard SSR, Stoichiometric amount of La$_2$O$_3$, SrCO$_3$ and Mn$_2$O$_3$ were mixed and ground for 10 hours then heated at 1200$^o$C for 20 hours. Planetary ball milling was used for nanostructuring the prepared material from SSR. For sol gel route, La(NO$_3$)$_3$.6H$_2$O, Sr(CH$_3$COO)$_2$ and Mn(CH$_3$COO)$_2$ were dissolved in deionized  water in stoichiometric amount. All the individual water solutions were mixed and required amount of citric acid (C$_6$H$_8$O$_7$) was added. The mixed solution was then heated at 120$^o$C till gel formation and then at 150$^o$C until the entire solution turns into black powder. The powder were then ground for 30 minutes and heated at 900$^o$C for 2 hours. All the prepared samples were characterized for phase purity by X-ray Diffraction (XRD) using Cu-K$_alpha$ radiation ($\lambda = 1.5406\, \AA$). Particle size distributions were calculated by using Scherer’s formula. The surface morphology of all the samples was observed by Field Effect Scanning Electron Microscope (FESEM). The magnetic measurement of samples was completed using Vibration Sample Magnetometer (VSM) at room temperature in a magnetic field of ±2\,T.  

\section{Results and Discussion}
LSMO samples prepared by different routes are of single phase, show rhombohedral structure with space group R$-3$c as shown in figure 1(a). It has been observed that the full width at half maximum (FWHM) of peak increases as we go from SSR to sol gel to Ball milling route. The FWHM is inversely proportional to the crystallite size of the sample, according to Scherer’s formula, which is given by
						
\begin{equation}
d=\frac{K \lambda}{\beta \cos \theta}
\end{equation}

Here $K$ ~ 0.9 (constant) and $\lambda$ = wavelength of X-ray. 
Figure.~2. shows the FESEM images of LSMO samples. The reduction in crystallite size has been confirmed by FESEM images
Figure.~3. shows the change in coercive field (H$_C$) and saturation magnetization (M$_S$) as a function of FWHM ($\beta$). It has been observed that the M$_s$ and remnant magnetization (M$_r$) is maximum and H$_c$ is minimum for samples having large crystallite size (i.e. for SSR) while as the crystallite size decreases, H$_c$ increases and M$_r$ and M$_s$ decreases.

The magnetization in the materials are affected by domain walls, formation of which depends on balance of several energies \cite{ref2} like exchange energy which is been used to align the magnetic moment of the atoms in the material in the same direction while orientation of magnetic moments in a particular direction is due to magneto crystalline anisotropy and elimination of magnetization of the material is due to magneto static energy. When the sample consists of large crystallite size, they consist of multi domain structures in which the magnetization regions are separated by domain walls. These domain walls result due to balance of the external magneto static energy and domain wall energy. The former increases as the volume of the particle increases and latter is due to materials having large interfacial area between the domains. As the size of the crystallite decreases, the volume reaches a critical value called critical volume, below which the more energy is required to create a domain wall than to balance the external magneto static energy. 

The values of M$_s$, M$_r$ and H$_c$ for different synthesis route are shown in table.~1. It has been observed that the reduced particle size also leads to a higher coercivity. When the particle size are reduced enough to have only single domain particle and all the spins in the domain are aligned in the same direction and since there is no domain wall therefore when we apply a magnetic field in opposite direction to demagnetize the sample, sample gets demagnetized via spin rotation rather than domain wall motion \cite{ref3}. This leads to high coercivity of the low crystallite size materials. Also the decrease in magnetization is attributed to the fact that as the particle size of the material decreases, spins which are aligned in a particular direction gets disordered due to strain created at the grain boundary due to small crystallite size of the material and therefore the saturation magnetization of the material reduces. 

\section{Conclusion}
We have successfully synthesized the LSMO samples through different synthesis route. Crystallite size peak broadening effect has been confirmed by FESEM. The magnetic properties of the samples suggest that as the crystallite size is reduced to a critical value, the disordered spins in the materials at the grain boundary leads to low magnetization as well as spin rotation demagnetization leads to high coercive field in the samples.

\newpage
	
\begin{table}
%\begin{sidewaystable}
\centering
\caption[Magnetic parameters as a function of synthesis route]{Magnetic parameters as a function of synthesis route}
\begin{tabular}{| c | c | c | c |}
\hline
Synthesis Route & Saturation Magnetization (emu/g) & Remnant Magnetization (emu/g) & Coercive Field (Oe)\\
\hline
Solid State Route & 60.77 & 55.30 & 20\\
\hline
Sol Gel Route & 40.60 & 31.60 & 859\\
\hline
Ball Milling & 33.06 & 23.84 & 1283\\
\hline
\end{tabular}
\end{table}

\newpage

\begin{figure} %fig.1 
\includegraphics[height=13.0 cm]{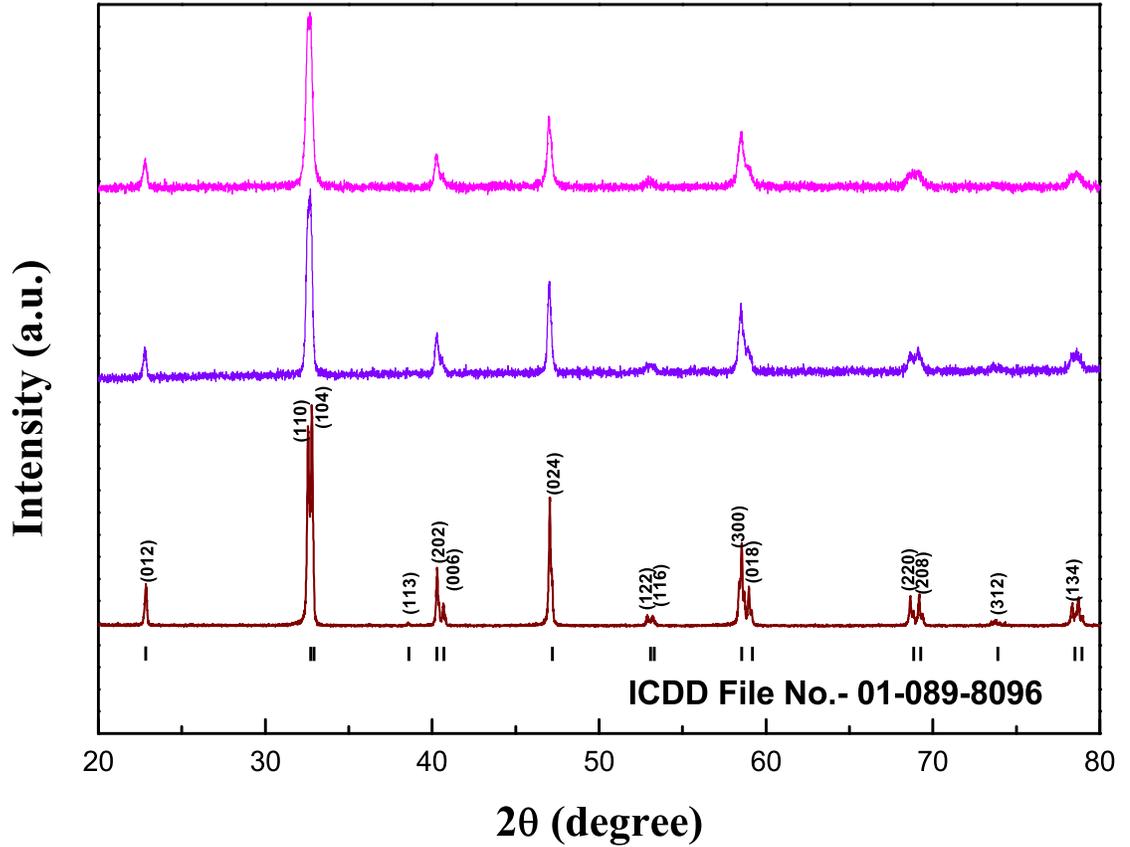}
\caption{(Color online) (a) XRD pattern of La$_{0.7}$Sr$_{0.3}$MnO$_3$ samples prepared by (i) Solid State Route (ii) Sol Gel route (iii) Ball Milling. (b) XRD peak broadening of La$_{0.7}$Sr$_{0.3}$MnO$_3$ samples for (i) Solid State Route (ii) Sol Gel route (iii) Ball Milling.}
	\end{figure}
\begin{figure} %fig.2
\includegraphics[height=3.6 cm]{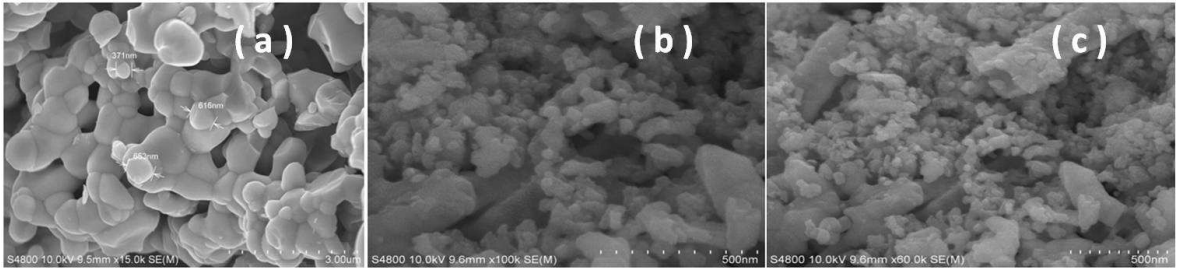}
\caption{(Color online) FESEM images of $LSMO$ samples prepared by (a) Solid State Route (b) Sol Gel route (c) Ball Milling.} 
\end{figure}
\begin{figure} %fig.3
\includegraphics[height=12.0 cm]{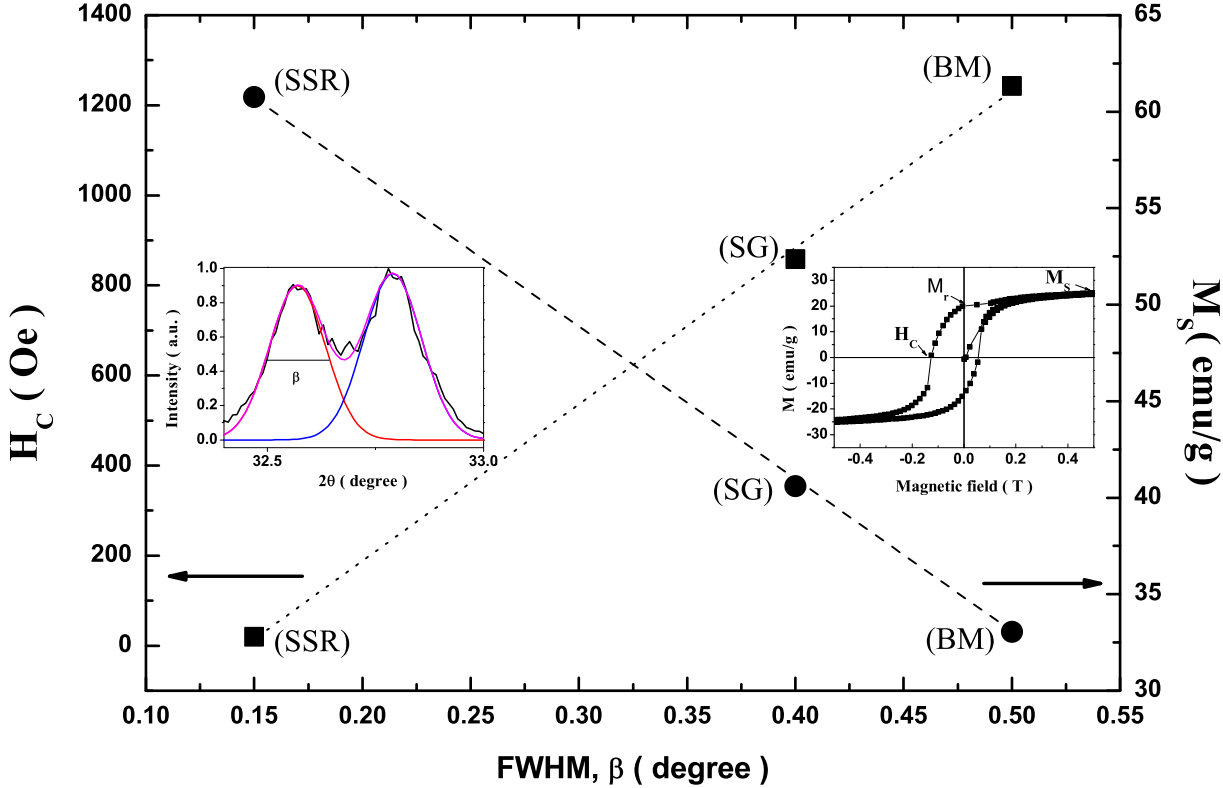}
\caption{(Color online) Variation of coercive field (H$_C$) and saturation magnetization (M$_S$) as a function of FWHM (for SSR- 0.15$^o$, for SG- 0.40$^o$ and for BM- 0.50$^o$). Left inset shows the FWHM found by fitting the XRD data. Right inset is the M-H curve of $LSMO$ samples prepared by BM route, showing the M$_S$, M$_r$ and H$_C$.}
\end{figure}

\end{document}